\documentclass[aps,prl,twocolumn,groupedaddress,amsmath,amssymb]{revtex4}

\usepackage{graphicx}
\usepackage{dcolumn}
\usepackage{bm}

\begin{document}

\title{$\pi$-Electron Spectrum of Oligoperylene[$N\times\cal N$]
-- Graphene Macromolecule and Its Derivatives}

\author{Alexander Onipko}
\email[]{aleon@bitp.kiev.ua}
\author{Lyuba Malysheva}
\affiliation{Bogolyubov Institute for Theoretical Physics, 03680 Kyiv, Ukraine}

\date{\today}

\begin{abstract}
We revisit the model of graphene as a $\pi$-conjugated macromolecule \cite{PRL} and show that it allows to understand the origin and describe, in many cases explicitly, the electronic structure of graphene sheets, achiral graphene ribbons and carbon tubes of any size, and also  graphene tori on the basis of the exact solution to the eigenvalue problem for the respective structures. The exact expressions of dispersion relations and detectable observables for these structures are compared with the predictions of other related models demonstrating that the revisited model either suggests the exact versions of widely used approximations or shows their limited applicability. 
\end{abstract}


\maketitle

{\bf Introduction}

In the tsunami wave of theoretical works on graphene dominated numerical calculations. Most of a comparatively small number of analytical studies in one way or another are based on the standard one-parameter tight-binding model of 2D-graphite launched by Philip R. Wallace in 1947 \cite{Wallace}. Since then, the heuristic power of this and related analytical descriptions of the electronic properties of graphene-based structures has been convincingly documented soon after discovery of carbon tubes \cite{Saito}. 

Except graphene macromolecule model (GMM), all earlier suggested analytical models can be applied to one class of graphene structures but are inappropriate to another despite the latter is a close relative to the former. For this reason, it is difficult or impossible to compare the scattered over dozens of publications predictions, which have been made regarding closely related species such as pairs of relatives: armchair graphene ribbons -- zigzag carbon tubes (aGR--zCT) and zigzag graphene ribbons -- armchair carbon tubes (zGR--aCT). As an example, in an early breakthrough theoretical study \cite{Saito} followed by tens if not hundreds of related publications, the electronic structures and their classification elaborated for aCT and zCT families of carbon tubes do not fit the related graphene ribbons which were studied with the use of approximate models \cite{Ando,BrF,Beenakker,Gunlycke,Review}. 

\begin{figure}[b!]
\includegraphics[width=0.45\textwidth]{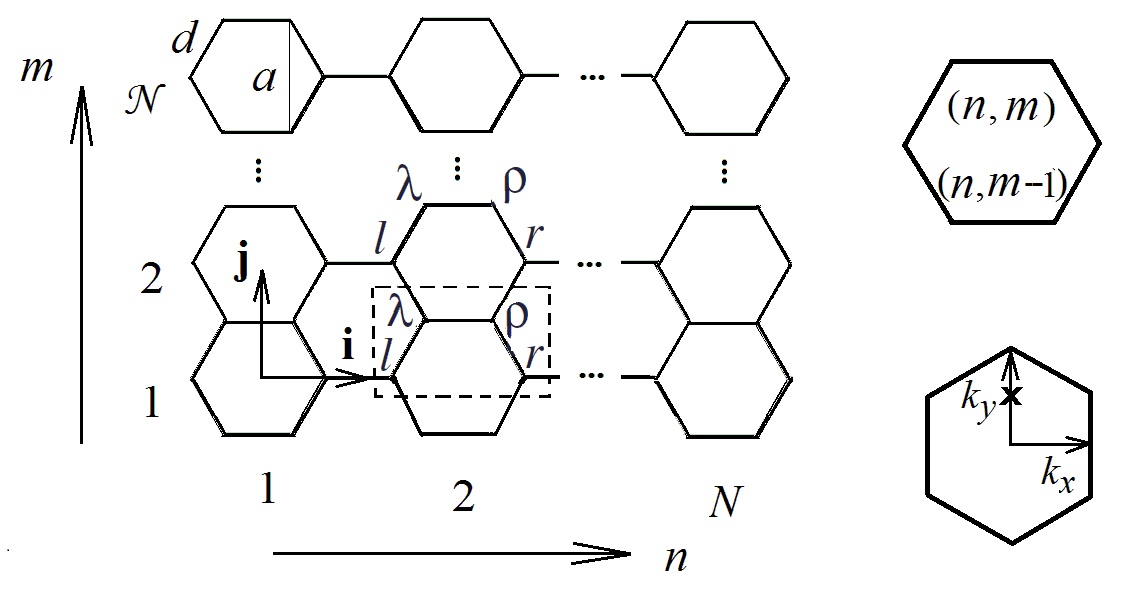}
\caption{Clarification of C-atom labeling in a $\sqrt{3}aN$-wide and $a\cal N$-long $\pi$-conjugated oligoperylene.
Dashed frame indicates the lattice elementary cell in the corresponding
$N$$\times$${\cal N}$ honeycomb lattice; $a$ and $d$ -- the minimal translation distance and the C--C bond length, respectively. A hexagon in 
the reciprocal k-space shows the specific point ${\bf k}^{\text x}=(0,2\pi/3)$ of the dispersion relation for armchair carbon tubes and tori explained in the text.}\label{Fig1}
\end{figure}

This limitation is not inherited by GMM sketched in Fig.~\ref{Fig1}. It is relevant to a thinkable plane macromolecule oligoperylene[$N\times\cal N$] \cite{Jay}. It is just ($\sqrt{3}aN$-long) oligoperylene if $\cal N$=2. Otherwise, it is understood here as graphene macromolecule of the size $\sqrt{3}aN$ in {\bf i} direction, $ a{\cal N}$ in {\bf j} direction, where $N$ and $\cal N$ are any integers. The advantages, as well as further applications of this model are concisely discussed in this Letter.

\begin{table*}[t!]
\caption{Dispersion relations for achiral graphene ribbons, 
carbon tubes, and tori}
\begin{tabular}{clcc}
  \\
aGTOR,aCT&\qquad $E^\pm_\nu(k)= \sqrt{1
\pm 4\left |\cos\dfrac{\pi \nu }{N}\right|\cos\dfrac{k}{2} +
4\cos^2\dfrac{k}{2}},$  &
$\begin{array}{c}
0\leq k\leq \pi,\\[5pt]
\nu=0,1,\dots,N-1. 
\end{array}$ & \qquad (T1)
\\
\\
zGTOR,zCT&\qquad $E^\pm_j(k)=\sqrt{1\pm 4\cos\dfrac{\sqrt{3}k}{2}
\left|\cos\dfrac{\pi j}{{\cal N}}\right| +
4\cos^2\dfrac{\pi j}{{\cal N}}},$ & $\begin{array}{c}
0\leq \sqrt{3}k\leq \pi,\\[5pt]
j=0,1,\dots,{\cal N}-1.
\end{array}$ &\qquad  (T2)\\
\\
 aGR  &\qquad $E^\pm_j(k)= \sqrt{1
\pm 4\cos\dfrac{\sqrt{3}k}{2}\cos\dfrac{\pi j}{2({\cal N}+1)} +
4\cos^2\dfrac{\pi j}{2({\cal N}+1)}},$  &
$\begin{array}{c}
0\leq \sqrt{3}k\leq \pi, \; {\bf k}\parallel {\bf i}   \\[5pt]
j=1,\dots,{\cal N},
\end{array}$ &\qquad  (T3)
 \\  
  \\
zGR&\qquad $E_\nu^\pm(k)=\sqrt{1
\pm 4\left |\cos\dfrac{\sqrt{3}k^\nu }{2}\right|\cos\dfrac{k}{2} +
4\cos^2\dfrac{k}{2}},$ & 
$\begin{array}{c}
0\leq k\leq \pi,\; {\bf k}\parallel {\bf j} \\[5pt]
\nu=0,1,\dots,N-1,
\end{array}$
&\qquad (T4a)
 \\&& & \\
&\qquad $\dfrac{ \sin(\sqrt{3}k^\nu N) }
{\sin\left[\sqrt{3}k^\nu\left(N+\dfrac{1}{2}\right)\right]} =\mp
2\cos\dfrac{k}{2},  $ &&\qquad (T4b)
\end{tabular}\label{Table1}
\end{table*}

{\bf Basic Equations}

As shown in \cite{PRL,PRB}, the full spectrum of $\pi$-electrons in the $[N,{\cal N}]$ all-carbon honeycomb rectangular lattice framed by armchair and zigzag edges is described by a pair of equations
\vspace{0mm}
\begin{align}\label{1}
&E_j^\pm(k^\nu_x) \notag \\
 &= \sqrt{1
\pm 4\left|\cos\frac{\sqrt{3}k^\nu_x }{2} \right|
\cos\frac{\pi j}{2({\cal N}+1)} +
4\cos^2\frac{\pi j}{2({\cal N}+1)}},
\end{align}
\begin{equation}\label{2}
\frac{ \sin(\sqrt{3}k^\nu_xN) }
{\sin\left[\sqrt{3}k^\nu_x\left(N+\frac{1}{2}\right)\right]} =\mp
2\cos\frac{\pi j}{2({\cal N}+1)},
\end{equation}
where the energies
$0\le E_j^-(k^\nu_x) \le \sqrt{5}$, $1\le E_j^+(k^\nu_x) \le 3$ are in units of the absolute value of the C--C hopping integral, 
and index  $\nu = 0,1,...,N\!-\!1$ 
numbers $N$ solutions to Eq. (\ref{2}) for each value of integer $j = 1,2,...,\cal N $. Here and in what follows, only the conduction band is in focus because of the spectrum symmetry with respect to the mirror reflection
in the
[$E_j^\pm(k^\nu_x)=E_{F}=0$]-plane, $E_{F}$ is the Fermi energy. 

For an integer value of $j =2({\cal N}+1)/3 \equiv j^*$, Eq. (\ref{1}) simplifies to
\begin{equation}\label{3}
E^+_{j^*}(k^\nu_x) = 2\cos\frac{\sqrt{3}k^\nu_x}{4}, \quad
E^-_{j^*}(k^\nu_x) = 2\sin\frac{\sqrt{3}k^\nu_x}{4}.
\end{equation} 

The only but not minor difficulty of the suggested description  is that the classification of electronic states requires the knowledge of solutions to the transcendental equation (\ref{2}). The latter is not far-off relative of the Lennard-Jones equation \cite{LJ} that refers to $\pi$-conjugated $N$-long polyenes and has $N$ solutions. Equation (\ref{2}) also has $N$ solutions but for each value of $j$. These can be either real or imaginary solutions which correspond respectively to either extended or decaying (local) states. An immediate conclusion is that the parentage of the local states in rectangular graphene sheets, armchair ribbons and carbon tubes with zigzag-shaped termini, and zGRs is one and the same, namely, zigzag edges. The existence of the local zigzag edge states was observed for the first time in the output of {\it ab initio} calculations \cite{Kobayashi}.

\begin{figure}[b!]
\includegraphics[width=0.48\textwidth]{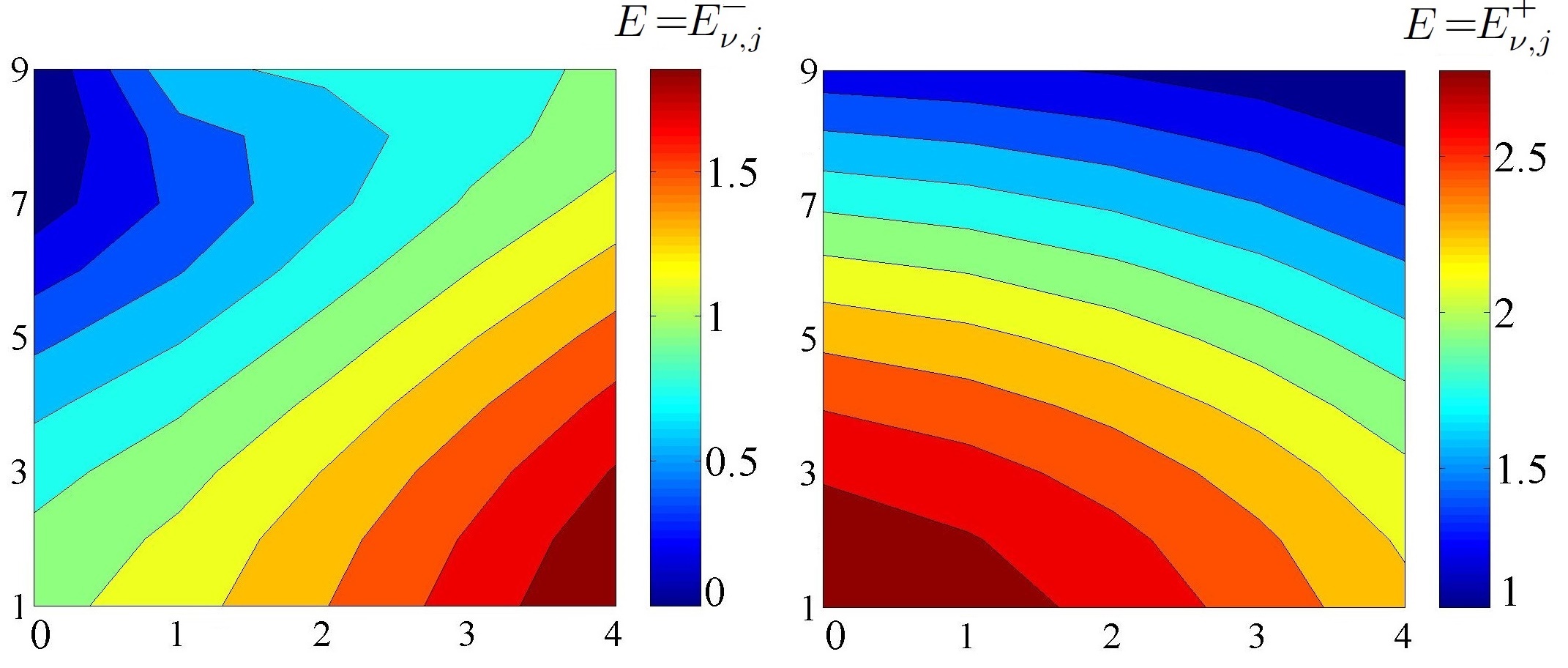}
\caption{
Equipotential diagrams for 5$\times$9  (almost square) graphene sheet:
$E=E^-_{\nu, j}$ -- left, $E=E^+_{\nu, j}$ -- right.   $\nu$- and $j$-values are indicated on the $x$ and 
$y$ axis, respectively.  Energy-color relationship is 
different for $E=E^-_{\nu,j}$ and $E=E_{\nu,j}^+$. The change in color corresponds to the lines of equal potential.}
\label{Fig2}
\end{figure}

Recent finding of highly accurate explicit expressions for solutions to 
Eq. (\ref{2}) \cite{Roots} removes the above mentioned difficulty almost completely. This result in combination 
with Eqs. (\ref{1}) and (\ref{2}) provides the fully analytical description of the 
$\pi$-electron spectrum and states in the rectangular fragment of graphene lattice. A particular example of such spectrum, which clarifies the interrelation between the eigenenergies $E_{\nu,j}^\pm$ and $\nu$, $j$ quantum numbers is shown in Fig.~\ref{Fig2}.

Equations (\ref{1}) and (\ref{2}) were derived with the Dirichlet or open boundary conditions. If instead we use the periodic boundary conditions (PBC), we obtain 
\begin{equation}\label{4}
E^\pm_{\nu,j} = \sqrt{1
\pm 4\left|\cos\frac{\pi \nu}{N}\cos\frac{\pi j}{\cal N}\right| +
4\cos^2\frac{\pi j }{\cal N}},
\end{equation}
where $\nu = 0,1,\dots, N-1$, $j = 0,1,\dots,{\cal N}-1$. Furthermore, the replacements
of $\pi\nu/{N}$ with $\sqrt{3}k_x/2$ and $\pi j/{\cal N}$ with $k_y/2$, $0\le k_x,k_y\le 2\pi$ converts Eq. (\ref{4}) into a counterpart to the Wallace equation describing the zone structure of 2D graphite,
\begin{equation}\label{5}
E^{\pm}(k_x,k_y) = \sqrt{1
\pm 4\left|\cos\frac{k_x}{2}\cos\frac{k_y}{2}\right| +
4\cos^2\frac{k_y}{2}}.
\end{equation}

{\bf Electronic Structure of Basic Graphene Derivatives}

{\it Finite-Length Achiral Graphene Ribbons.}--As such equations (\ref{1}) and (\ref{2}) 
are straightforwardly applicable to $\sqrt{3}aN$-long and 
$a{\cal N}$-wide armchair graphene ribbon -- aGR[N,${\cal N}$] and also to $a{\cal N}$-long and $\sqrt{3}aN$-wide zigzag 
graphene ribbon -- zGR[N,${\cal N}$].

{\it Finite-Length Achiral Carbon Tubes.}--With the  replacement of $\sqrt{3}k^\nu_x$ by $2\pi\nu/N$ 
(the use of PBC in $\bf i$ direction), Eq. (\ref{1}) determines the spectrum of
$a{\cal N}$-long armchair carbon tube with $N$ hexagons along the 
circumference (aCT[N,${\cal N}$]). The replacement of $\pi j/[2({\cal 
N}+1)]$ with $\pi j/{\cal N}$ in Eqs. (\ref{1}) and (\ref{2}) (the use of PBC in $\bf j$ direction) makes them relevant to the description of the 
$\sqrt{3}a N$-long zigzag carbon tube with ${\cal N}$ hexagons 
along the circumference referred here as zCT[N,${\cal N}$].\\

\begin{figure*}[ht!]
\includegraphics[width=0.97\textwidth]{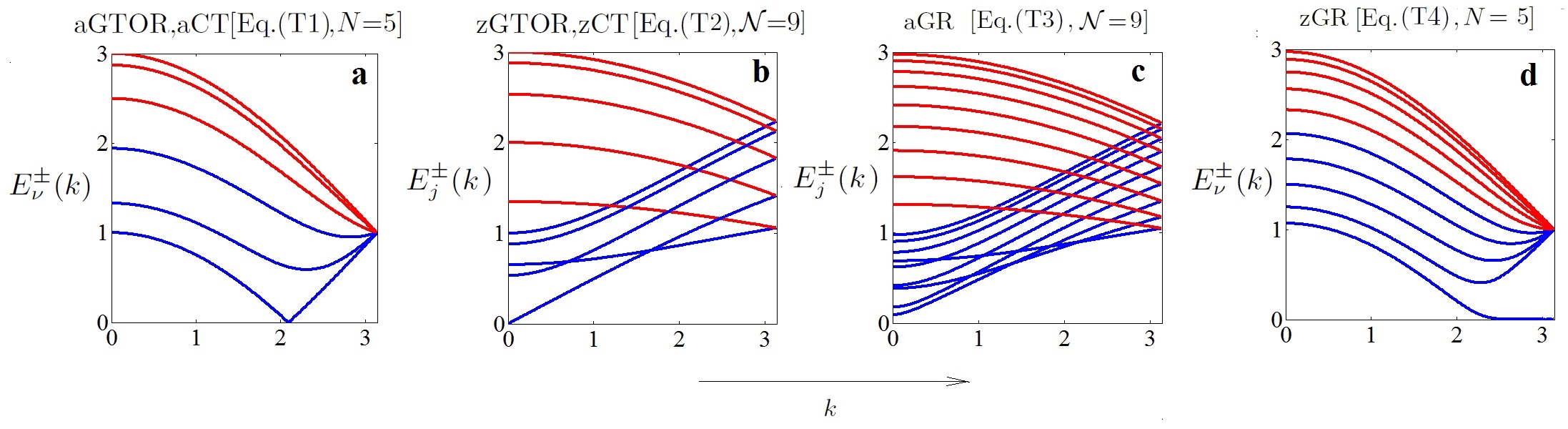}
\caption{Conduction band spectrum {\normalsize{\bf a}} for aGTOR and aCT, 
the $\nu$th dispersion curve labeling
according to Eq. (T1): $E^-_0(k)$, $E^-_1(k)=E^-_4(k)$, $E^-_2(k)=E^-_3(k)$,
$E^+_2(k)=E^+_3(k)$, $E^+_1(k)=E^+_4(k)$, $E^+_0(k)$;
 {\normalsize{\bf b}} for zGTOR and zCT, the $j$th dispersion curve labeling
according to Eq. (T2): $E^-_3(0)=E^-_6(0)$, $E^-_2(0)=E^-_7(0)$, $E^-_4(0)=E^-_5(0)$, 
$E^-_1(0)=E^-_8(0)$, $E^-_0(0)$, $E_j^+(0)=E_{9-j}^+(0)$, $j$=4,3,2,1, $E^+_0$;
{\normalsize{\bf c}} for aGR, the $j$th dispersion curve labeling
according to Eq. (T3): $E^-_7(k)$, $E^-_6(k)$,$E^-_8(k)$, $E^-_5(k)$, $E^-_4(k)$, $E^-_9(k)$, $E^-_3(k)$,  $E^-_2(k)$, $E^-_1(k)$, $E^+_j(k)$, $j=9,8,...,1$;
{\normalsize{\bf d}} for zGR, the $\nu$th dispersion curve $E_j^\pm(k)$ labeling according to Eq. (T4): 
$E^-_\nu(0)$, $\nu=0,1,2,3,4$, $E^+_\nu(0)$, $\nu=4,3,2,1,0$.
In all panels, $E^-_{\nu(j)}$-curves are blue, $E^+_{\nu(j)}$-curves are red; labeling $E_\nu^\pm(k)$ and  $E_j^\pm(0)$ is from down to up.}
\label{Fig3}
\end{figure*}

{\bf Dispersion Relations}

{\it Achiral Graphene Tori.}--This family of  graphene structures can be thought as a result of uniform bending of aCT[$N,{\cal N}$] 
until carbon atoms on one end of the tube  map their counterparts on the opposite end:
$[n,1,\lambda] = [n,{\cal N}+1,\lambda]$ and
$[n,1,\rho] = [n,{\cal N}+1,\rho]$. 
Under the condition ${\cal N}>>N$ such that ensures negligible distortions of C--C bonds, Eq. (\ref{4}) determines the \underline{$\nu$th mode} dispersion relation for a torus with armchair pattern on its surface (aGTOR). Similarly, under  the condition $N>>{\cal N}$, Eq. (\ref{4}) can be regarded as the \underline{$j$th mode} dispersion relation for the torus with zigzag pattern on its surface (zGTOR). Thereby, $2\pi j/{\cal N}\equiv k$ and $2\pi\nu/N\equiv k$ in the cases of $\text{aGTOR}$ and $\text{zGTOR}$ respectively, have the meaning of 1D quasi-continuous wave vector along the torus perimeter. The dispersion relations for $\text{aGTOR}$ and $\text{zGTOR}$ appear in Table~\ref{Table1} as Eqs. (T1) and (T2), respectively.

{\it Achiral Graphene Ribbons aGRs and zGRs.}--In the limiting case $N$$\rightarrow$$\infty$, Eq. (\ref{1}) gives the exact dispersion relation for the aGR family of graphene ribbons, Eq. (T3). In the limit ${\cal N}$$\rightarrow\infty$, Eqs. (\ref{1}) and (\ref{2}) give the exact dispersion relation for zGR family of graphene ribbons  specified in Eqs. (T4a) and (T4b).

{\it Achiral Carbon Tubes aCTs and zCTs.}--The replacement of $2\pi j/{\cal N}$ with $k$ in Eq. (\ref{4}) gives the exact dispersion relations for armchair carbon tubes in the form of Eq. (T1). The only difference from the aGTOR dispersion relation is that in the given case ${\bf k}\parallel{\bf j}$. Likewise, the replacement of $2\pi\nu/N$ by $k$ in Eq. (\ref{4}) leads to the exact dispersion relations for zigzag carbon tubes in the form of Eq. (T2). In the latter case, ${\bf k}\parallel{\bf i}$.

The dispersion relations for aGR and zGR determine the electron kinematics in the honeycomb lattice in twelve directions along zigzag and armchair pathways. The change of the direction by the angle of $\pi/3$ radians results in the change of the pathway character from zigzag-like to armchair-like or {\it vise versa} and hence, in the change of electron kinematics. The dispersion relations for achiral graphene ribbons and carbon tubes are closely interconnected. This relationship near the Fermi level and its manifestation in the conductance quantization has been examined in detail in \cite{ZhEThF}.

{\bf Discussion}

The comparison of aCT and zCT spectra shown in Figs.~\ref{Fig3}a and ~\ref{Fig3}b with the results of pioneering paper \cite{Saito} shows the point to point coincidence. Notice, the Saito et al. electronic portrait of aCT reproduced here is in a sharp controversy with the dispersion curves (or bands) shown in Fig. 9a \cite{White2005}.

The aGTOR and zGTOR spectra as well as the aCT and zCT spectra differ from each other even qualitatively, Figs.~\ref{Fig3}a,b. This is not at all surprising in view of the substantial distinctions between the respective dispersion relations (T1) and (T2). In particular, the position of the band minima and the interband spacing are determined by different equations. For aCT, it follows from Eq. (T1) that 
min$\left \{E_{\nu}^-(k)\right\}$ = $\sin(\pi\nu/N)$. 
In the case of zCT, Eq. (T2) gives 
min$\left\{E_{j}^-(k)\right\}$ = 
$\left| 1-2\left |\cos(\pi j /{\cal N})\right|\right|$. 

In contrast, the zGR spectrum in Fig.~\ref{Fig3}d  has a lot in common with the aCT spectrum in Fig.~\ref{Fig3}a. The observed difference between the number of dispersion curves was expected. Disregarding the two-fold spin degeneracy, all the $N_\nu=10$ bands in the zGR-spectrum are nondegenerate. But the respective ($\nu$$\neq$$0$)-bands in the aCT-spectrum are two-fold degenerate. The same is true for the pair aGR-zCT. Obviously, the source of the $\pi$-electron states degeneracy in the aCT and 
zCT spectra is connected with PBC. 

The similarity of dispersion curves in Figs.~\ref{Fig3}a and ~\ref{Fig3}d breaks for the ($\nu$=0)th mode. In Fig.~\ref{Fig3}a, $E^-_0(k)
=\left| 1-2\cos(k/2)\right |$, whereas in Fig.~\ref{Fig3}d, the behavior of $E^-_0(q)$ ($q\equiv k-2\pi/3$) is qualitatively different if $q < q^c \le\pi$, where
\begin{equation}\label{6}
q^c\equiv 2\arccos\frac{N}{2N+1}. 
\end{equation}
Under this condition, Eq. (T4b) has only one imaginary solution for each value of $q$. It can be shown that for $N\gtrsim 10$, the larger part of the zero-mode dispersion $E_0^-(q)$ (which represents the edge-states dispersion) is governed by two exponential dependencies, 
$E_0^-(q)$=$\dfrac{\sqrt{3}}{2}\exp(-\sqrt{3}qN)$ and 
$E_0^-(q)$=$\left(\pi/3-q \right)^{2N}$ when $q$ approaches 
$\pi/3$. The crossover between the two is within the interval 
$\pi/12<q<\pi/6$. With the increase of zGR width 
$q^c$$\rightarrow$$2\pi/3$ so that the energy of edge states 
$E_0^-(k)$ becomes practically zero within the interval 
$2\pi/3<k<\pi$. Such a detailed quantitative analytic description of the really unique, exponential-like dispersion of zigzag-edge states  has never been reported before.

 \begin{table}[t!]
 \caption{The lowest electron energy in achiral graphene tori, ribbons, and carbon  tubes $E_{\text min} $. Band gap 
$E_{\text g}$ = 2$E_{\text min}$, $R_{a(z)}$ denotes a(z)CT radius, $W_{a(z)}$ -- a(z)GR width.}
 \begin{tabular}{cccc}
 \\
Structure&\quad Metallic & \quad Semicond &\;CT-Radius   \\ 
& & &GR-Width   \\[7pt]
 aGTOR, aCT&\quad 0 &&  R$_a$=$\sqrt{3}Na/(2\pi)$ \\[7pt]
zGTOR, zCT &\quad 0\footnote{if $\cal N$ is divisible by 3} &
 $\dfrac{\pi}{\sqrt{3}{{\cal N}}}$
  & R$_z$=${\cal N}a/(2\pi)$  \\[7pt]
 aGR &\quad 0\footnote{if ${\cal N}+1$ is divisible by 3} &
 $\dfrac{\pi}{\sqrt{12}{({\cal N}+1)}}$
 & W$_a$=${\cal N}a$ \\[7pt]
  zGR&&$\quad \left (\dfrac{\pi}{{\cal N}+1} \right )^{2N}$  &
 W$_z$=$\sqrt{3} Na$   
 \end{tabular}\label{Table2}
 \end{table}
  
However, the conclusion just made concerns only {\it infinitely} long zigzag graphene ribbons which do not exist in real world. The band gap definition for zGR given in Table~\ref{Table2} is obtained for zGR$[N,{\cal N}]$ from the exact solution to Eqs. (1) and (2) under the condition ${\cal N}>>N$. It shows that for finite-width, finite-length zGRs, the band gap for the actual values of the length and width is not zero. Thus, zigzag graphene ribbons are not 2D metals. Instead, they are to be considered as semiconductors with the band gap highly sensitive to changes of zGR length $a\cal N$, or length $\sqrt{3}aN$, or both. The zGR electronic structure near the Fermi level obtained in the {\bf k}$\cdot${\bf p} approximation \cite{BrF} has little to do with the above presented results which follow from the exact solution of the spectral problem.  

{\bf Concluding remarks}

We emphasize that the Wallace equation which describes the zone structure of 2D graphite was obtained with the only purpose to classify the parallelogram-like honeycomb lattice in the canonical way of Solid State theory that requires the minimal number of inequivalent atoms (two in the case of graphite) in the elementary cell plus the use of PBC when solving the eigenvalue problem. Literally, the 2D graphite lattice subjected to PBC, as well as its counterparts achiral TORi, do not correspond to any real properly defined object which has been observed and studied experimentally. Therefore, the use of 2D graphite zone structure, the more so, the approximations originating from this model may well lead to partly or completely misleading results. A few of such results have been briefly mentioned in the above discussion.

Long-standing goal to synthesize ordered, carbon based 2D macromolecules is claimed to be becoming the reality \cite{Nature2}. However, as before it is a long way to go \cite{Nature1}. At this stage of development, it is of crucial importance to have at hand a controllable, exactly solvable, and realistic model aimed to describe analytically the electronic spectra of 2D $\pi$-conjugated graphene macromolecule and its derivatives. It is shown here that such model, the only one reported thus far, can be expressed in the form of two exact equations (1) and (2) for graphene, while their modifications provide the description of the most extensively studied graphene daughter structures. In one way or another, this instrumental pair of equations has been used for the comprehensive analysis of some problems such as, e.g., electron transmission through graphene-based structures \cite{PRB12a,PSS15} but a large number of others appeal to be addressed on the rigorous basis.


\begin{thebibliography}{99}
\bibitem{Jay}
The chemical name of graphene macromolecule has been suggested by Dr. Jaywant Phopase.
\end{thebibliography}
\end{document}